\documentclass[aps,a4paper]{revtex4}

\usepackage{graphicx}
\usepackage{bm,amsmath,amssymb}
\usepackage{natbib}

\providecommand\upi{\pi}
\providecommand\bnabla{\nabla}
\providecommand\bcdot{\cdot}

\newcommand\Ku{\mbox{\textit{Ku}}}  

\newcommand\etal{\mbox{\textit{et al.}}}

\begin{document}

\title{Weak--strong clustering transition in renewing
compressible flows}

\author{Ajinkya Dhanagare, Stefano Musacchio and Dario Vincenzi}
\affiliation{Universit\'e Nice Sophia Antipolis, CNRS, Laboratoire J. A. Dieudonn\'e,  
UMR 7351, 06100 Nice, France}

\begin{abstract}
We investigate the statistical properties of Lagrangian tracers transported 
by a time-correlated compressible renewing flow.
We show that  
the preferential sampling of the phase space performed by tracers 
yields significant differences between 
the Lagrangian statistics and its Eulerian counterpart. 
In particular, the effective compressibility experienced by tracers
has a non-trivial dependence on the time correlation of the flow. 
We examine the consequence of this phenomenon on the clustering of tracers, 
focusing on the transition from the weak- to the strong-clustering regime. 
We find that the critical compressibility at which the transition occurs
is minimum when the time correlation of the flow is of the order of 
the typical eddy turnover time. 
Further, we demonstrate   
that the clustering properties in time-correlated compressible flows are 
non-universal 
and are strongly influenced by the spatio-temporal structure of the velocity field. 
\end{abstract}

\hfill{\textit{J. Fluid Mech.}, in press}

\maketitle

\section{Introduction}

The dynamics of tracers in turbulent flows has important
applications in a variety of physical phenomena ranging from the 
dispersion of atmospheric pollutants \citep{C73}
to the transport of plankton in the oceans \citep{A98}. 
Moreover, the motion of tracers is intimately related to
the mixing properties of turbulent flows and therefore
determines the statistics of passive fields such as
temperature in a weakly heated fluid or
the concentration of a dye in a liquid \citep{FGV01}. 
The last fifteen years have seen a renewed interest in
the Lagrangian study of turbulence thanks to
the development of new experimental and numerical particle-tracking
techniques \citep{TB09,PPR09}.

Several of the qualitative properties of tracer dynamics
in turbulent flows have been understood by means of the
\citet{K68} model, in which the velocity is a homogeneous and isotropic
Gaussian field with zero correlation time and power-law spatial
correlations. Under these assumptions, the separations between tracers
form a multi-dimensional diffusion process with space-dependent diffusivity,
the properties of which have been studied analytically \citep{FGV01,CFG08}.
In particular, in the smooth and incompressible regime of the Kraichnan model,
tracers behave chaotically and
spread out evenly in the fluid. If the velocity field is
weakly compressible, 
the Lagrangian dynamics remains chaotic, but
tracers cluster over a
fractal set with Lyapunov dimension $1<D_L<d$, where $d$ is the
spatial dimension of the fluid \citep{LJ84,LJ85,CKV98}. 
The Lyapunov dimension
decreases as the degree of compressibility increases;
furthermore,
the density of tracers exhibits a multifractal behaviour in space,
which indicates the presence of strong fluctuations in the distribution
of tracers within the fluid \citep{BGH04}.
Finally, if the degree of compressibility 
of the Kraichnan model exceeds a critical value,
all the Lyapunov exponents
of the flow become negative and tracers collapse onto a pointlike fractal
with $D_L=0$.
This latter regime only exists if $d\leqslant 4$ \citep*{CKV98} and
is known as the regime of strong compressibility.

In nature, compressible flows are found not only for large values of the Mach
number. An example of a low-Mach-number compressible flow is given by
the velocity field on the free surface of
a three-dimensional incompressible flow \citep{SO93,CDGS04}.
Furthermore, at small Stokes numbers, the dynamics of inertial
particles in an incompressible flow can be assimilated to
that of tracers in an effective compressible velocity field \citep{M87,BCDM07}.
Compressible flows like those mentioned above 
have a nonzero correlation time.
In this respect, the Kraichnan model is not realistic and quantitative
discrepancies may be expected between the theoretical predictions
and the experimental and numerical observations.
In a numerical simulation of a turbulent surface flow,
\citet{BDES04} have found that $D_L$ does
decrease as a function of the degree of compressibility, in accordance
with the prediction of the Kraichnan model. However, 
the transition from the weak- to the strong-clustering
regime occurs at a larger degree of compressibility compared to
the Kraichnan model.
This phenomenon is counterintuitive, because the level of clustering
would be expected to increase when the correlation of the flow is nonzero.
Thus, the study by \cite{BDES04} raises the questions of the interplay
between compressibility and temporal correlation in turbulent flows
and of the degree of universality of the weak--strong clustering transition
\citep[for further studies
on clustering in turbulent surface flows, see][]{SE02,BDDL06,VFF07,DP08,LG10,
LGB10,LMS13,PM14}. 
The investigation of the universality of this phenomenon 
is particulartly interesting in the light of previous findings
of non-universal transport properties in random compressible flows
for passive scalar fields \citep{E00}.

For inertial particles, 
the effect of the temporal correlation of the flow on the Lagrangian dynamics
has been studied analytically in the following one-dimensional cases:
for a velocity gradient described by the telegraph noise \citep{F07} or
by the Ornstein--Uhlenbeck process \citep{W11} and 
for a velocity field given by a Gaussian potential 
with exponential correlations
\citep{GM13a}.
In the case of tracers in compressible flows, \citet{CGHKV03} have studied
two-particle dispersion in Gaussian self-similar random fields.
\citet{FMM07} have calculated the Lyapunov exponent and the 
statistics of the stretching rates for a one-dimensional
strain described by the telegraph noise.
\citet{GM13b} have obtained the Lyapunov exponents of a two-dimensional
random flow in the limits of short and long correlation times; 
the solenoidal and potential components of the velocity 
were assumed to be Gaussian random functions with exponential
spatio-temporal correlations.

In this paper, we undertake a thorough study of the effects of 
temporal correlations on tracer dynamics 
in a compressible random flow. 
We consider a compressible version of the two-dimensional renewing flow,
which consists of a random sequence of sinusoidal velocity profiles
with variable origin and orientation. Each profile remains
frozen for a fixed time; 
by changing the duration of the frozen phase, we can vary the correlation
time of the flow and examine the effect on clustering.
The renewing flow, in its original incompressible version, 
has been successfully applied to the study of the kinematic
dynamo \citep{Z84,GB92}, of chaotic mixing \citep{P94,A96,Y99,E00,TDG04,AT11}, of
inertial-particles dynamics \citep{E02,P12}
and of polymer stretching \citep{MV11}.
The properties of this model flow allow us to fully characterise the
Lagrangian statistics of tracers as a function of the degree of compressibility
and for a wide range of correlation times.
We show that, in a time-correlated
compressible flow, even single-time
Lagrangian averages can differ considerably from their Eulerian
counterparts. Furthermore, we demonstrate 
that the properties of clustering do not depend only on universal
parameters such as the degree of compressibility and the Kubo number, but
also on the specific spatial and temporal properties of the velocity field. 
In particular, 
we show that a crucial role is played by the spatial distribution 
of the stagnation points.

The rest of the paper is organized as follows. In \S~\ref{section:renewing},
we introduce the compressible renewing flow and describe its Eulerian
properties. In \S~\ref{sec:lagr-euler}, we compare the Lagrangian and
Eulerian statistics of tracer dynamics as a function of the degree of
compressibility and of the correlation time of the flow.
Section~\ref{section:fractal} describes the fractal clustering of tracers
in the weakly compressible regime and the weak--strong clustering transition.
Section~\ref{section:conclusions} concludes the paper by discussing
the non-universal character of the weak--strong clustering transition.

\begin{figure}
\centering
\includegraphics[width=0.34\textwidth]{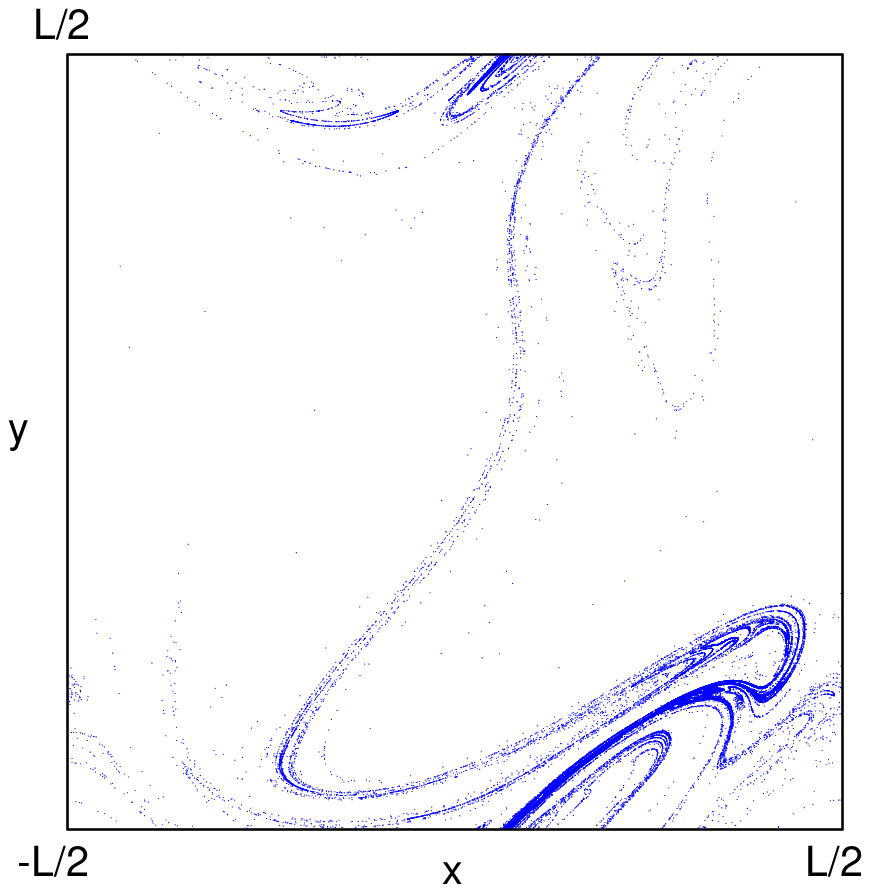}%
\hspace{7mm}%
\includegraphics[width=0.34\textwidth]{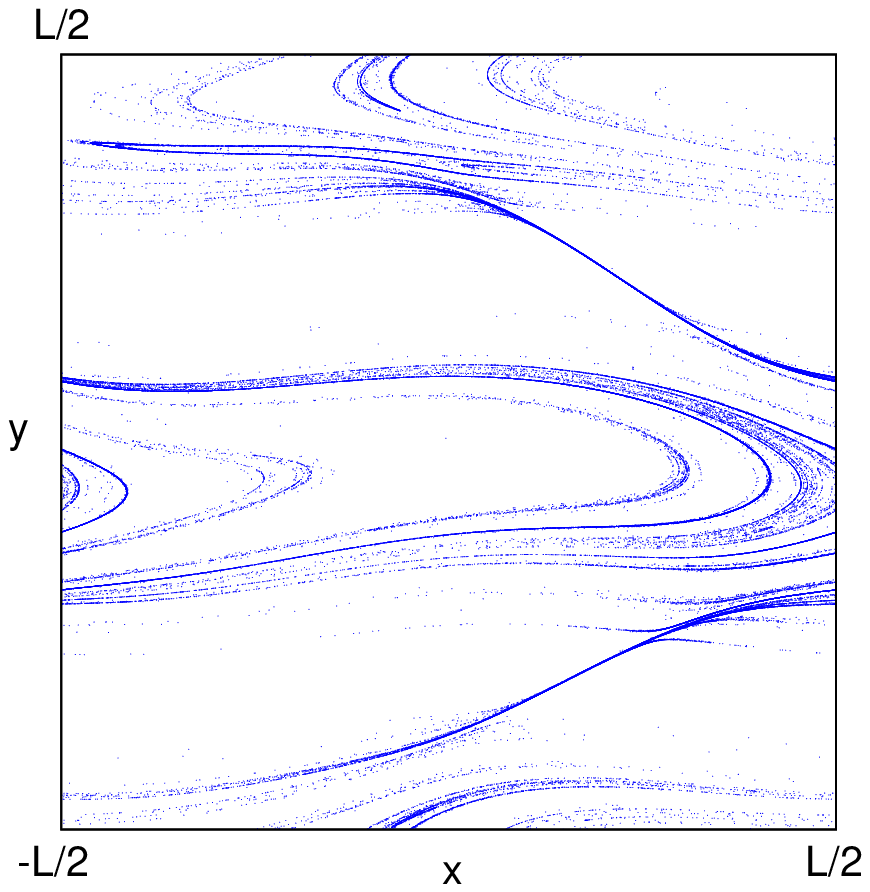}%
\caption{(Colour online) Spatial distribution of tracers in the compressible
renewing flow for $ C=1/4$ and $\Ku=0.1$ (left), $\Ku=10$ (right). 
Each plot shows the positions of $10^5$ tracers.
For $\Ku=10$, the distribution of tracers mirrors the structure of the
flow, which consists of periodic channels and of regions where transport is
inhibited by the stagnation lines.}
\label{fig:snapshot}
\end{figure}

\section{Compressible renewing flow}
\label{section:renewing}

We consider the following velocity field $\bm u=(u_x,u_y)$
on a periodic square box $\varOmega=[-L/2,L/2]^2$:
\begin{equation}
\label{eq:flow-1}
\begin{cases}
u_x=U\sqrt{2(1- C)}\cos(ky+\phi_y)+U\sqrt{2 C}\cos(kx+\phi_x),
\\
u_y=0,
\end{cases}\quad 2nT\leqslant t<(2n+1)T
\end{equation}
and
\begin{equation}
\label{eq:flow-2}
\begin{cases}
u_x=0,
\\
u_y=U\sqrt{2(1- C)}\cos(kx+\phi_x)+U\sqrt{2 C}\cos(ky+\phi_y),
\end{cases}\quad (2n+1)T\leqslant t<2(n+1)T,
\end{equation}
where $n\in\mathbb{N}$,
$k=2\upi/L$, $U=\sqrt{\langle u^2\rangle}$ is the root-mean-square velocity,
and $ C=\langle(\bnabla\cdot\bm u)^2\rangle/\langle%
\Vert\nabla\bm u\Vert^2_F\rangle$ is the degree of compressibility of the flow.
Here, $\Vert\cdot\Vert_F$ is the Frobenius norm and
$\langle\cdot\rangle$ denotes a spatial average over the domain $\varOmega$. 
Note that $0\leqslant C\leqslant 1$;
$C=0$ corresponds to an incompressible flow, whereas
$ C=1$ corresponds to a gradient flow.
The angles $\phi_x$ and $\phi_y$ are independent random numbers uniformly 
distributed over $[0,2\upi]$ and change randomly at each time
period $T$.

The velocity field defined in \eqref{eq:flow-1} and \eqref{eq:flow-2}
is a sequence of randomly translated sinusoidal profiles, 
each of which persists for
a time $T$; the velocity is alternatively oriented in the $x$ and
$y$ directions.
The renewing flow is in principle non-stationary, because the values
of the velocity at two different times 
are either correlated or independent depending on 
whether or not the two times belong to the same frozen phase.
However, it can be regarded as stationary for times much longer than $T$
\citep{Z84}.
The temporal statistics of the flow is characterised by
the correlation function:
\begin{equation}
\label{eq:correlation}
F_{\mathcal{E}}(t)\equiv \dfrac{\langle \bm u(\bm x,s+t)\bcdot\bm u(\bm x,s)\rangle_{\mathcal{E}}}%
{U^2}=
\begin{cases}
1-\dfrac{t}{T}, & t\leqslant T,
\\
0 & t>T,
\end{cases}
\end{equation}
where $\langle\cdot\rangle_{\mathcal{E}}$ denotes a space-time Eulerian
average: $\langle\cdot\rangle_{\mathcal{E}}\equiv T^{-1}L^{-2}
\int_0^{T}\int_\varOmega\cdot\;\mathrm{d}s\mathrm{d\bm x}$.
The correlation time of the flow is:
$T_{\mathcal{E}}\equiv\int_0^T F_{\mathcal{E}}
(t)\mathrm{d}t=T/2$.
A dimensionless measure of the correlation time is given by the
Kubo number $\Ku\equiv TUk$; $\mathit{Ku}$ is proportional to
the ratio of $T_{\mathcal{E}}$ and the eddy turnover time
of the flow.

\begin{figure}
\centering
\includegraphics[width=0.33\textwidth]{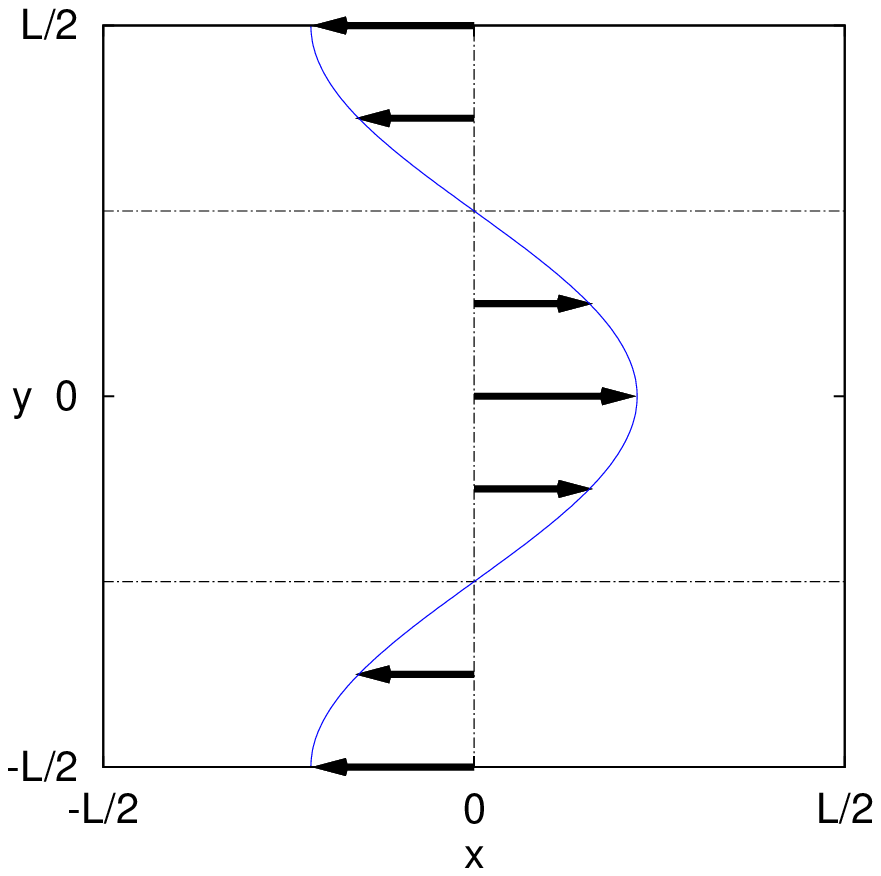}\hfill%
\includegraphics[width=0.33\textwidth]{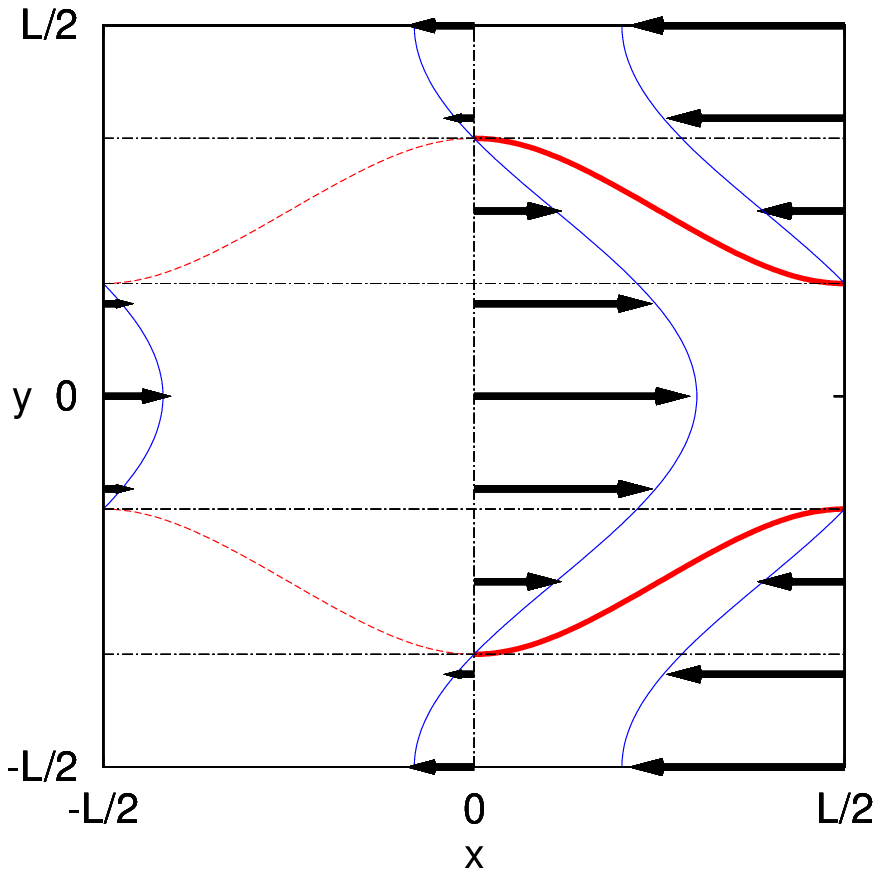}\hfill%
\includegraphics[width=0.33\textwidth]{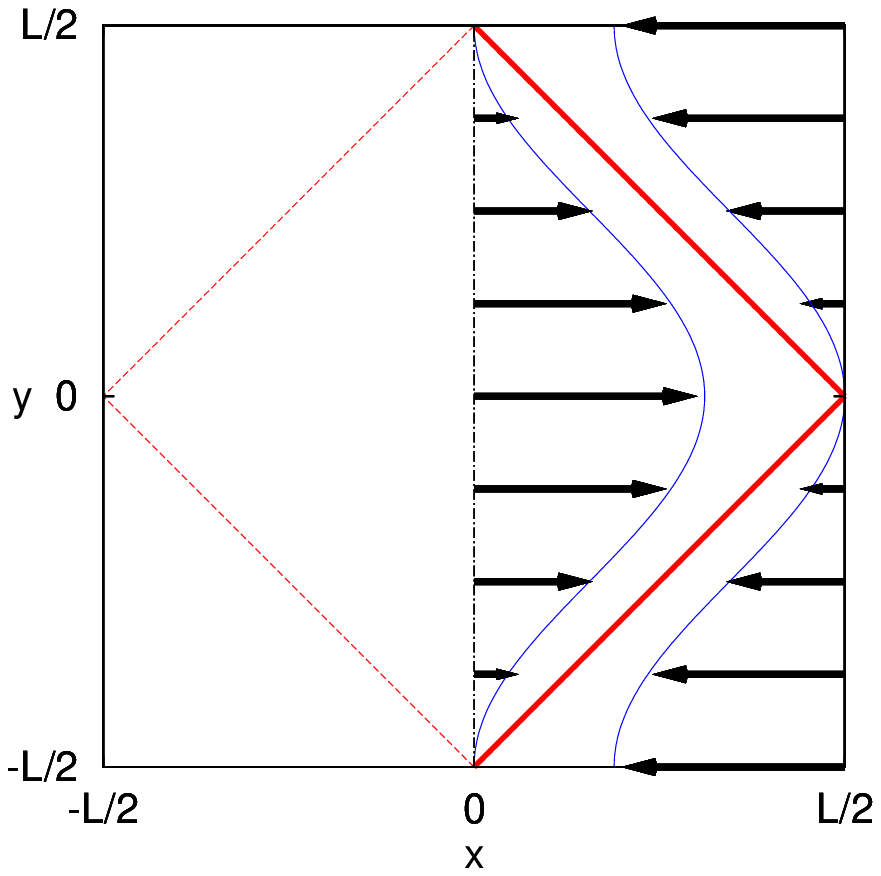}
\caption{(Colour online) Velocity profile
of the renewing flow for $ C=0$ (left), $ C=1/4$ (middle)
and $ C=1/2$ (right). The solid red lines are the portions
of the stagnation lines where $\nabla\cdot\bm u<0$.
}
\label{fig:stagnation}
\end{figure}

The position of a tracer evolves according to the following equation:
\begin{equation}
\label{eq:trajectory}
\dot{\bm X}(t)=\bm u(\bm X(t),t).
\end{equation}
Figure \ref{fig:snapshot} shows the distribution of tracers
in the weakly compressible regime. 
The transition from the regime of weak compressibility to that
of strong compressibility occurs when the maximum Lyapunov exponent
of the flow becomes negative. In this case, the flow is not chaotic
anymore and tracers are attracted to a pointlike set.
To study this transition,
it is useful to consider
the set of stagnation points of the flow, 
in which $\bm u=(0,0)$. 
Tracers indeed tend to accumulate in the neighbourhood
of these points.
From \eqref{eq:flow-1} and \eqref{eq:flow-2},
we deduce that
the set of stagnation points of the renewing flow consists of the two lines:
\begin{equation}
\label{eq:stagnation-1}
y=\pm\dfrac{1}{k}\arccos\bigg(\sqrt{\dfrac{ C}{1- C}}\cos(kx+\phi_x)\bigg)-\dfrac{\phi_y}{k}+2\upi m,
\end{equation}
if the velocity is given by \eqref{eq:flow-1}, or
\begin{equation}
\label{eq:stagnation-2}
y=\pm\dfrac{1}{k}\arccos\bigg(\sqrt{\dfrac{1- C}{ C}}\cos(kx+\phi_x)\bigg)-\dfrac{\phi_y}{k}+2\upi m,
\end{equation}
if the velocity is given by \eqref{eq:flow-2}.
In \eqref{eq:stagnation-1} and \eqref{eq:stagnation-2}, $m\in\mathbb{Z}$ is such that $(x,y)\in\varOmega$.
The stagnation lines are shown in figure~\ref{fig:stagnation} for some
representative values of the degree of compressibility.
If $ C=0$, the flow consists of parallel periodic `channels' of
width $L/2$. If $0< C<1/2$, the stagnation lines form barriers that
block the motion of tracers 
over a portion of the domain whose size increases
as $ C$ approaches 1/2; the width of the periodic channels shrinks
accordingly. 
Finally, if $1/2\leqslant C\leqslant 1$,
the stagnation lines divide the domain into regions
that are not linked by any streamline.
For these values of $ C$, there are no periodic trajectories and
if $\mathit{Ku}$ is sufficiently large, all tracers collapse onto
the stagnation lines.
We conclude that the transition from the regime of weak clustering
to that of strong clustering
must occur for $ C\leqslant 1/2$. However, the critical
value of the degree of compressibility depends on $\mathit{Ku}$.

\section{Lagrangian versus Eulerian statistics}
\label{sec:lagr-euler}

The properties of the flow introduced in \S~\ref{section:renewing},
namely the root-mean-square velocity, the degree of compressibility
and the correlation function, are of an Eulerian nature. 
If the flow is compressible and has a nonzero correlation time, 
the Lagrangian counterparts of the aforementioned
quantities may be different. Indeed, tracers 
are attracted towards the stagnation points and hence
do not sample the phase space uniformly.

\begin{figure}
\centering
\begin{minipage}{0.49\textwidth}
\includegraphics[width=\textwidth]{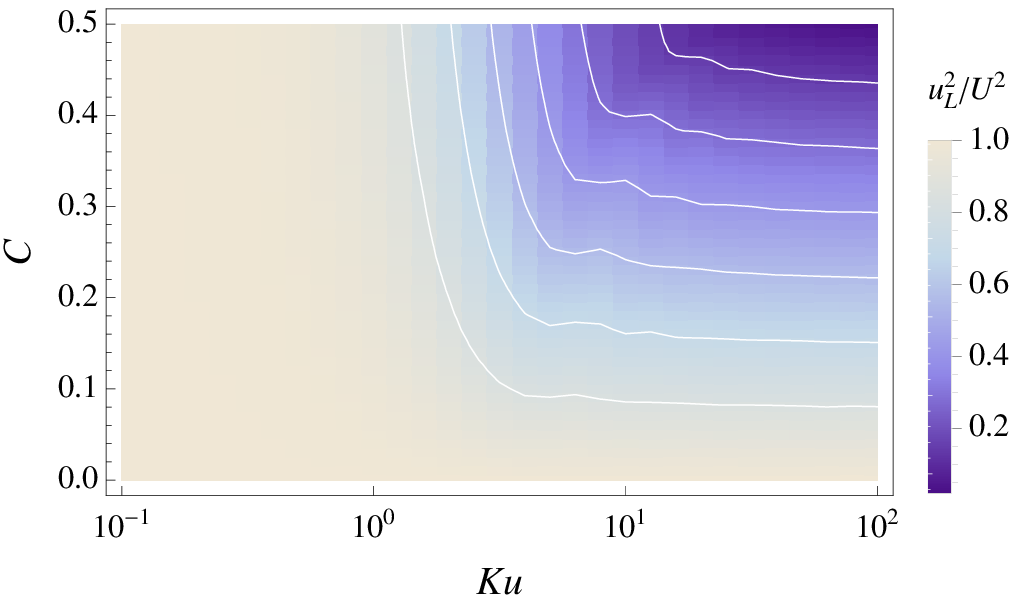}
\end{minipage}\hfill%
\begin{minipage}{0.49\textwidth}
\includegraphics[width=\textwidth]{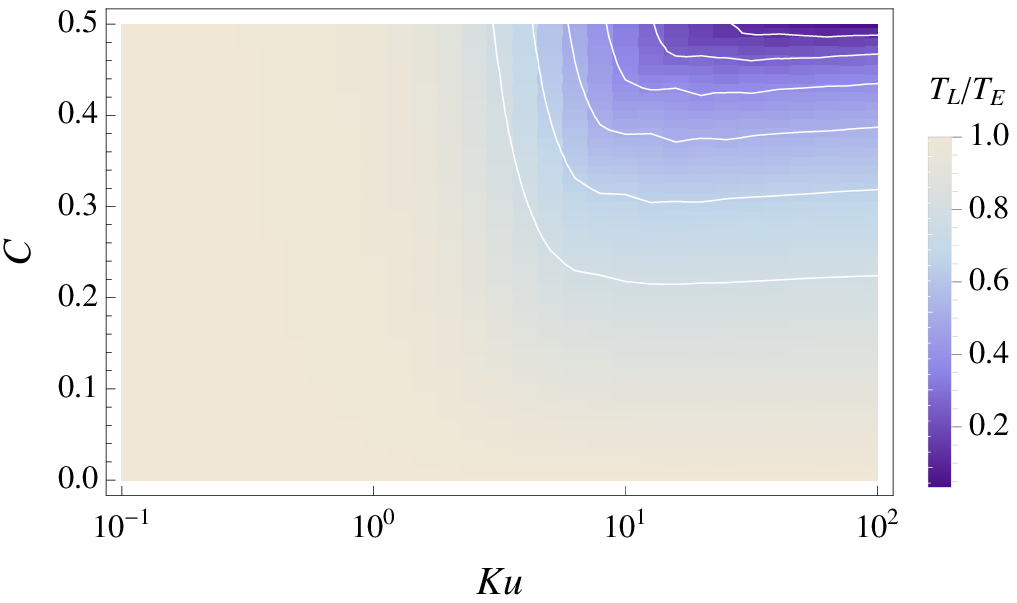}
\end{minipage}
\caption{(Colour online)
Left: Ratio of the mean-square Lagrangian and Eulerian velocities
as a function of the Kubo number $\Ku$ and of the degree of compressibility $ C$ 
of the flow. 
Right: Ratio of the Lagrangian and Eulerian correlation times
of the velocity as a function of $ C$ and $\Ku$.
}
\label{fig:urms_tc}
\end{figure}

We define the root-mean-square Lagrangian velocity as:
$u_{\mathcal{L}}\equiv\sqrt{\langle u^2(\bm X(s),s)\rangle_{\mathcal{L}}}$,
where $u=\vert\bm u\vert$ and $\langle\cdot\rangle_{\mathcal{L}}$ is
a Lagrangian average over both the random trajectory $\bm X(s)$ and time.
Figure~\ref{fig:urms_tc} (left panel) compares $u^2_{\mathcal{L}}$
and its Eulerian counterpart $U^2$ for different values of $\mathit{Ku}$ and
$ C$. The plot includes 51 values of $ C$ between 0 and 1/2 and
31 values of $\mathit{Ku}$ ranging from $10^{-1}$ to $10^2$.
For each couple $(\Ku,C)$, we have
computed $u^2_{\mathcal{L}}$ by solving \eqref{eq:trajectory}
for $10^2$ tracers and for an integration time $t=2\times 10^3 T$
(to integrate \eqref{eq:trajectory}, we have used
a fourth-order Runge--Kutta method).
At $ C=0$, 
$u_{\mathcal{L}}$ is the same as $U$ for all values of $\mathit{Ku}$,
because tracers explore the phase space uniformly.
Likewise, at $\mathit{Ku}=0$ the Eulerian and the Lagrangian
statistics coincide independently of the value of $ C$.
By contrast, for $ C\neq 0$ and $\mathit{Ku}\neq 0$,
$u_{\mathcal{L}}<U$ because tracers are attracted towards the stagnation lines,
where $u=0$.
The ability of the stagnation points to trap tracers
strengthens with increasing $ C$ and $\Ku$; hence
$u_{\mathcal{L}}$ eventually approaches zero.

The Lagrangian correlation function of the velocity is defined as:
\begin{equation}
\label{eq:corr-lagr}
F_{\mathcal{L}}(t)\equiv 
\dfrac{\langle \bm u(\bm X(s+t),s+t)\bcdot
\bm u(\bm X(s),s)\rangle_{\mathcal{L}}}%
{u_{\mathcal{L}}^2}.
\end{equation}
The associate Lagrangian correlation time is:
$T_{\mathcal{L}}\equiv\int_0^T F_{\mathcal{L}}(t)\mathrm{d}t$.
\begin{figure}
\centering
\begin{minipage}{0.49\textwidth}
\includegraphics[width=\textwidth]{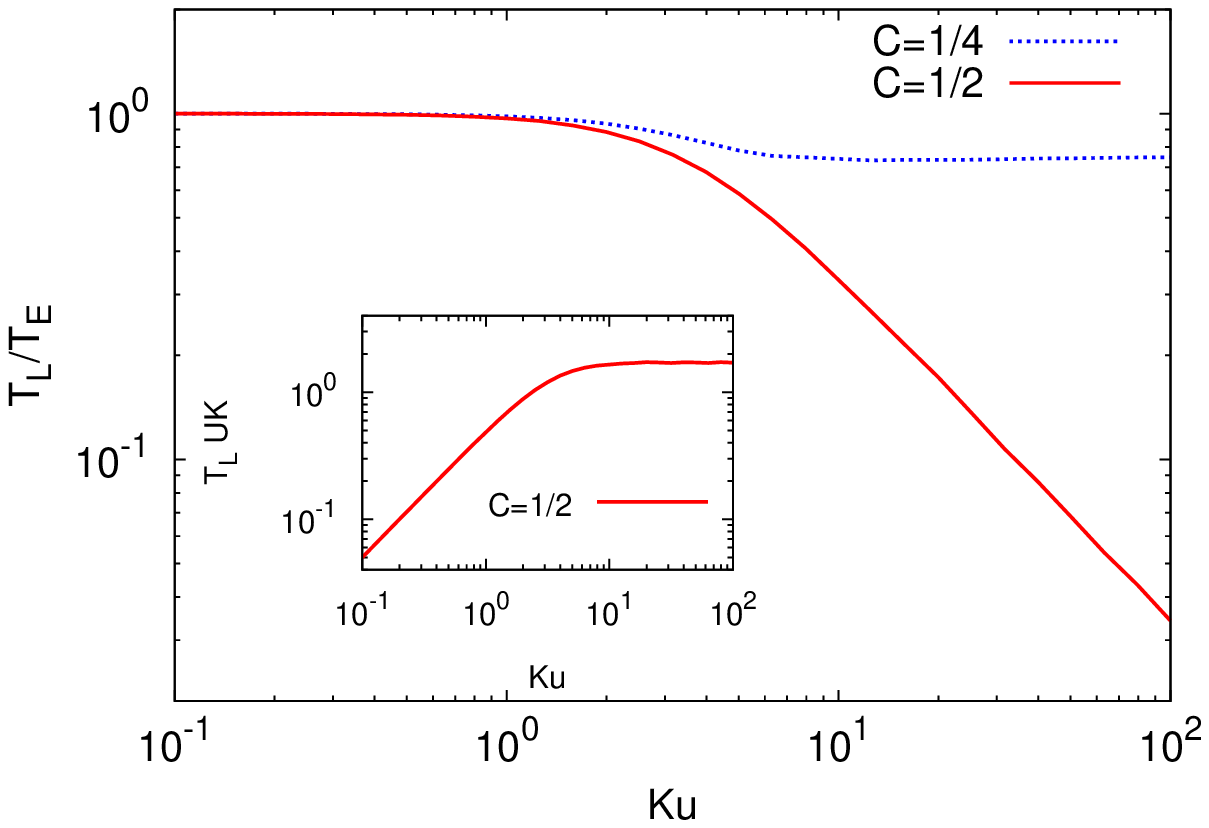}
\end{minipage}\hfill%
\begin{minipage}{0.49\textwidth}
\includegraphics[width=\textwidth]{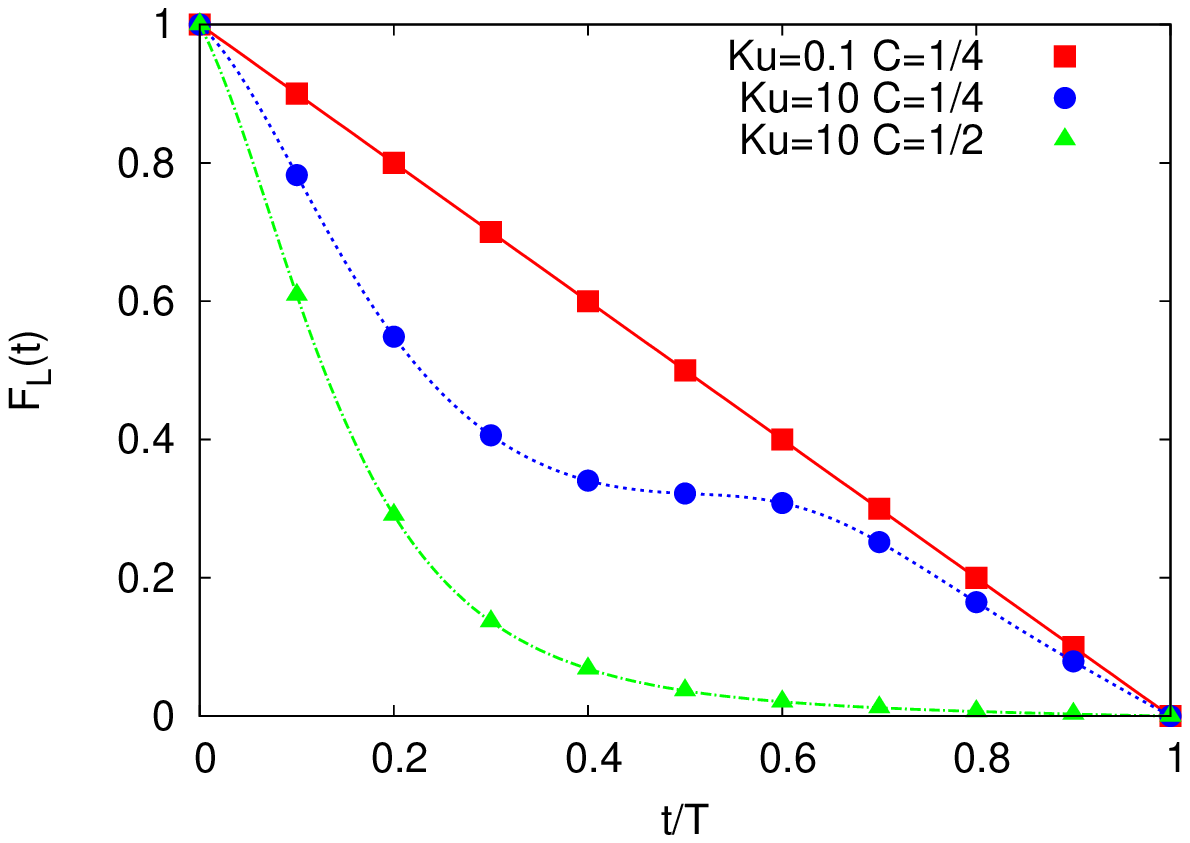}
\end{minipage}
\caption{(Colour online)
Left: Ratio of the Lagrangian and Eulerian correlation times
of the velocity as a function of $\Ku$ for $ C=1/4$ (dotted, blue curve)
and $ C=1/2$ (solid, red curve); the inset shows the 
Lagrangian correlation time rescaled by $(Uk)^{-1}$
as a function of $\Ku$ for $ C=1/2$.
Right: Lagrangian correlation function of the velocity
for $\Ku=0.1$, $ C=1/4$ (squares), 
$\Ku=10$, $ C=1/4$ (circles)
and $\Ku=10$, $ C=1/2$ (triangles). 
}
\label{fig:corr-time}
\end{figure}
If the flow is incompressible ($ C=0$) or 
if it is decorrelated in time ($\Ku=0$), 
then $T_{\mathcal{L}}=T_{\mathcal{E}}$ (figure~\ref{fig:urms_tc}, right panel).
If both $ C$ and $\Ku$ are nonzero, $T_{\mathcal{L}}$ is smaller than
$T_{\mathcal{E}}$ and the ratio $T_{\mathcal{L}}/T_{\mathcal{E}}$ decreases
as $\Ku$ or $ C$ increase (figure~\ref{fig:urms_tc}).
Again, this behaviour can be explained by considering 
that a tracer is attracted towards the stagnation points, where $u=0$,
and hence its velocity decorrelates from the velocity it had at the beginning of the period. 
As $\Ku$ and $ C$ increase, a larger fraction of tracers get close 
to the stagnation points, and therefore the decorrelation is faster.
The inset of figure~\ref{fig:corr-time} (left panel) also shows that,
for $C$ close to 1/2, 
$T_{\mathcal{L}}$ is proportional to $T$ for small values of $\Ku$, whereas
it saturates to a value proportional to 
the eddy turnover time for large values of $\Ku$. 
Indeed, after that time most of the tracers have reached a stagnation point, 
and their velocities have completely decorrelated. 
 

Figure~\ref{fig:corr-time} (right panel) shows that, for small values of $\Ku$,
$F_{\mathcal{L}}(t)$ is the same as $F_{\mathcal{E}}(t)$ irrespective of
the value of $ C$. However, at large $\Ku$, not only the Lagrangian correlation
time of the velocity, but also the functional form of
$F_{\mathcal{L}}(t)$ varies with $ C$.

The degree of compressibility experienced by tracers
also depends on $\Ku$ and $ C$ and differs from its Eulerian value
if $\Ku\neq 0$ and $ C\neq 0$.
Let us define the Lagrangian degree of compressibility as:
$C_{\mathcal{L}}\equiv
\langle(\bnabla\cdot\bm u)^2\rangle_{\mathcal{L}}/\langle%
\Vert\nabla\bm u\Vert^2_F\rangle_{\mathcal{L}}$. 
Then, 
$ C_{\mathcal{L}}> C$ for all nonzero values of $\Ku$ and $ C$,
because tracers spend more time in high-compressibility regions.
For fixed $\Ku$, the increase in compressibility is an increasing function of $ C$,
whereas for fixed $ C$ it is maximum 
when $\Ku$ is near to~1 and vanishes both in the small- and in the large-$\Ku$
limits (figure~\ref{fig:compressibility_lyap1}, left panel). 
The non-monotonic behaviour of the Lagrangian compressibility as a function of $\Ku$ 
is due to a peculiar feature of the model flow considered here. 
The stagnation lines 
\eqref{eq:stagnation-1} and \eqref{eq:stagnation-2}
toward which the tracers are attracted
do not coincide with the regions where the 
local compressibility of the flow is maximum, 
i.e., the lines $ky + \phi_y = m\upi$ for $2nT\leqslant t<(2n+1)T$ 
and $kx + \phi_x = m\upi$ for $(2n+1)T\leqslant t<2(n+1)T$.
Therefore, in the long-correlated limit 
the preferential sampling of the regions of strong 
compressibility is reduced.

\section{Fractal clustering}
\label{section:fractal}

The spatial distribution of tracers within a fluid
can be characterised in terms of the Lyapunov dimension
\citep[e.g.][]{SO93}:
\begin{equation}
D_L=N+\dfrac{\sum^N_{i=1}\lambda_i}{\vert\lambda_{N+1}\vert},
\end{equation}
where
$\lambda_1\geqslant\lambda_2\geqslant\cdots\geqslant\lambda_d$ are the Lyapunov
exponents of the flow and
$N$ is the maximum integer such that $\sum^N_{i=1}\lambda_i\geqslant 0$
($d$ denotes the dimension of the flow).
Three different regimes can be identified.
If the flow is incompressible ($\sum_{i=1}^d\lambda_i=0$), tracers spread out evenly
within the fluid ($D_L=d$). In the weakly compressible regime ($\sum_{i=1}^d\lambda_i<0$ and
$\lambda_1>0$), tracers cluster over a fractal set ($1<D_L<d$). 
In the strongly compressible
regime ($\sum_{i=1}^d\lambda_i<0$ and $\lambda_1<0$), tracers are attracted to a pointlike
set ($D_L=0$). 
The transition from the regime of weak compressibility
to that of strong compressibility occurs when $\lambda_1$ changes sign. 

\begin{figure}
\centering
\begin{minipage}{0.52\textwidth}
\includegraphics[width=\textwidth]{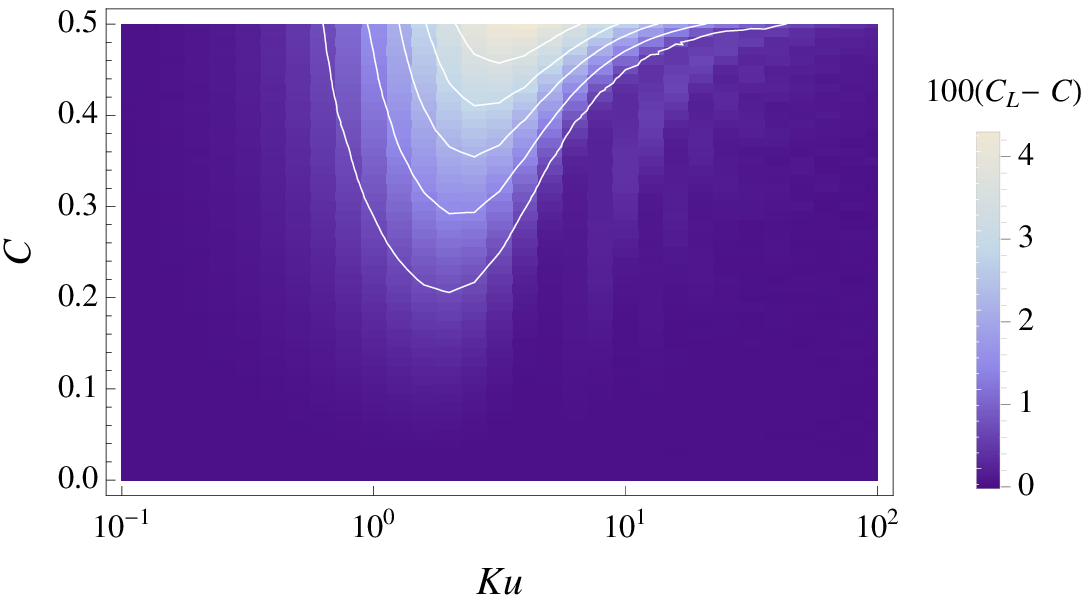}
\end{minipage}\hfill%
\begin{minipage}{0.48\textwidth}
\includegraphics[width=\textwidth]{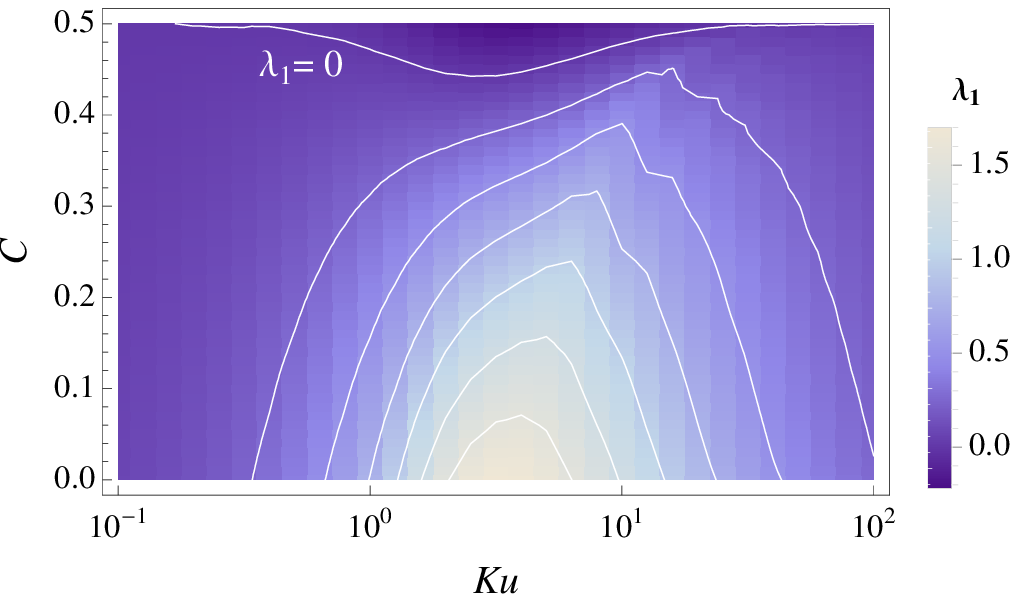}
\end{minipage}
\caption{(Colour online)
Left: Increase in compressibility multiplied by 100,
$(C_{\mathcal{L}}- C)\times 100$, as a function of
$\Ku$ and $ C$. 
Right:  
Maximum Lyapunov exponent of the compressible renewing flow
as a function of $\Ku$ and $ C$
}
\label{fig:compressibility_lyap1}
\end{figure}

For the smooth $d$-dimensional \citet{K68} flow, the Lyapunov exponents
can be calculated exactly (we remind the reader that in the Kraichnan model the velocity field is
Gaussian, delta-correlated in time, and statistically homogenous and isotropic).
The Lyapunov exponents are:
$\lambda_i=\mathcal{D}\{d(d-2i+1)-2 C[d+(d-2)i]\}$, where $i=1,\dots,d$,
$C$ is defined as in \S~\ref{section:renewing}, and 
$\mathcal{D}>0$ determines the
amplitude of the fluctuations of the velocity gradient \citep*{LJ84,LJ85}. 
Thus, for $d=2$,
the Lyapunov dimension of the smooth Kraichnan model is 
$D^0_L=2/(1+2 C)$ and the weak--strong clustering
transition occurs at $ C=1/2$.
In time-correlated flows, the prediction of the Kraichnan model is recovered in the small-$\Ku$ limit
\citep{BDES04,GM13b}.

In this section, we study how the weak--strong clustering transition depends on $\Ku$
in the compressible renewing flow.
To compute the Lyapunov exponents, we have used 
the method proposed by \cite{BGGS80}; we have set
the integration time to $t=10^6$ 
for all values of $\Ku$ and $ C$ in order
to ensure the convergence of the stretching rates to their asymptotic values.

The maximum Lyapunov exponent 
decreases with increasing $C$ (figure~\ref{fig:compressibility_lyap1}, right panel).
Its behaviour as a function of $\Ku$ is different in the weakly
compressible and in the strongly compressible regimes.
For small values of $ C$, $\lambda_1$ is maximum for
$\Ku$ near to~1, which signals an increase in chaoticity
when the correlation time of the flow is comparable to the eddy turnover time.
By contrast, for values of $ C$ near to 1/2, $\lambda_1$ is minimum
at $\Ku\approx 1$, in accordance with the fact that the 
Lagrangian compressibility
is maximum for these values of the parameters (figure~\ref{fig:compressibility_lyap1}).
We also note that if $\Ku$ is near to~1, $\lambda_1$ 
becomes negative
for $ C<1/2$; hence the weak--strong clustering transition 
occurs at a lower degree of
compressibility compared to the short-correlated case.

\begin{figure}
\centering
\begin{minipage}{0.495\textwidth}
\includegraphics[width=\textwidth]{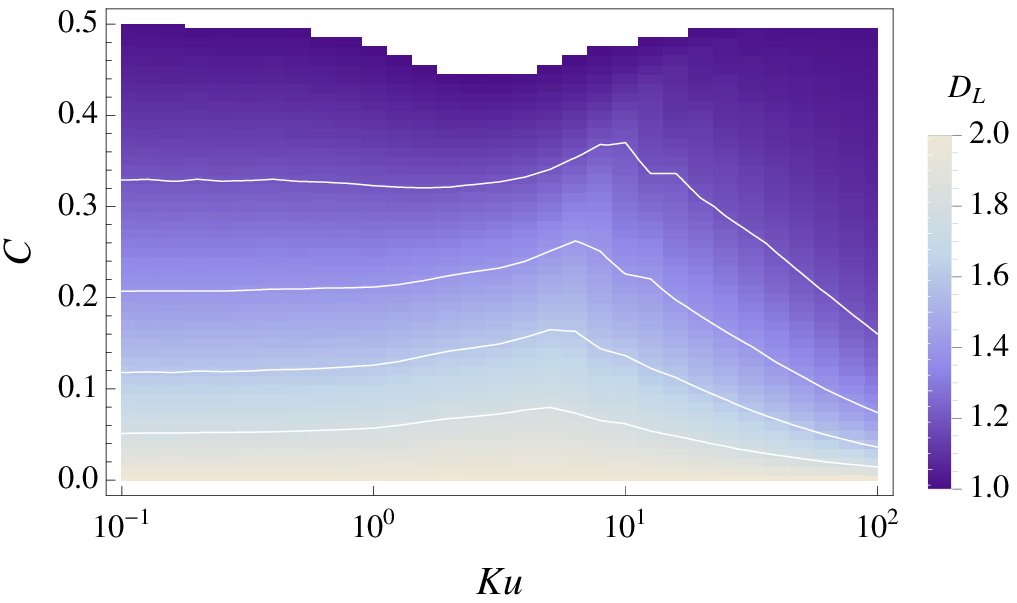}
\end{minipage}\hfill%
\begin{minipage}{0.505\textwidth}
\includegraphics[width=\textwidth]{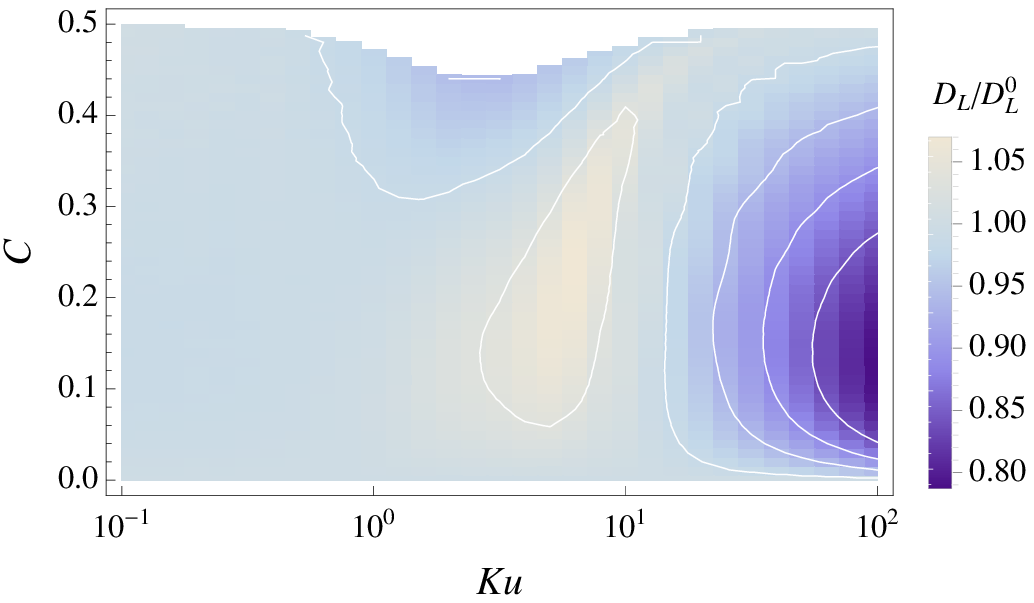}
\end{minipage}
\caption{(Colour online)
Left: Lyapunov dimension $D_L$ as a function of $\Ku$ and $ C$.
The region in which $D_L=0$ is couloured in white in the plot.
Right: $D_L$ rescaled by its value at $\Ku=0$ as a function of $\Ku$ and $ C$.}
\label{fig:D_L}
\end{figure}

Analogous conclusions can be reached by studying the behaviour of $D_L=1-\lambda_1/\lambda_2$ 
(figure~\ref{fig:D_L}).
For fixed $\Ku$, 
an increase in the Eulerian compressibility yields an increased level
of clustering. 
The behaviour as a function of $\Ku$ is not monotonic and depends on the
value of $C$. When $\Ku$ is near to 1, the level of clustering
is minimum if $C$ is small and is maximum if $C$ is near to 1/2;
furthermore,
$D_L$ vanishes for values of $ C$ smaller than the critical value of the short-correlated
case. The most important deviations of
$D_L$ from the $\Ku=0$ prediction are observed for values of $\Ku$ greater than 1
(figure~\ref{fig:D_L}, right panel).

\section{Conclusions}
\label{section:conclusions}

We have studied the Lagrangian dynamics of tracers in a 
time-correlated compressible random
flow as a function of the degree of compressibility and the Kubo number.
The use of the compressible renewing flow has allowed us to examine a wide
area of the parameter space $(\Ku,C)$. 
We have shown that, in compressible
random flows with nonzero correlation time, Lagrangian 
correlations differ significantly from their Eulerian counterparts, because
tracers are attracted towards the stagnation points and therefore do not
sample the phase space uniformly.
This fact influences the spatial distribution of tracers within the fluid.
In particular, 
in both the small- and the large-$\Ku$ limits,
the critical degree of
compressibility for the weak--strong clustering transition is the same as
for a short-correlated flow.
By contrast, when the correlation time of the flow is comparable to the eddy turnover time,
a smaller degree of compressibility is required for the transition to occur.
The non-monotonic behaviour of the critical degree of compressibility 
is a consequence
of the fact 
that the stagnation points do not coincide with the points in which the
compressibility is maximum.
This behaviour is very different from that observed by \citet{GM13b} in
a Gaussian velocity field with exponential spatio-temporal correlations. In
that flow, the critical degree of compressibility 
indeed decreseases monotonically as a function of $\Ku$ and 
tends to zero in the large-$\Ku$ limit, i.e. the weak--strong clustering
transition is more and more favoured as $\Ku$ increases.

\begin{figure}
\centering
\includegraphics[width=0.55\textwidth]{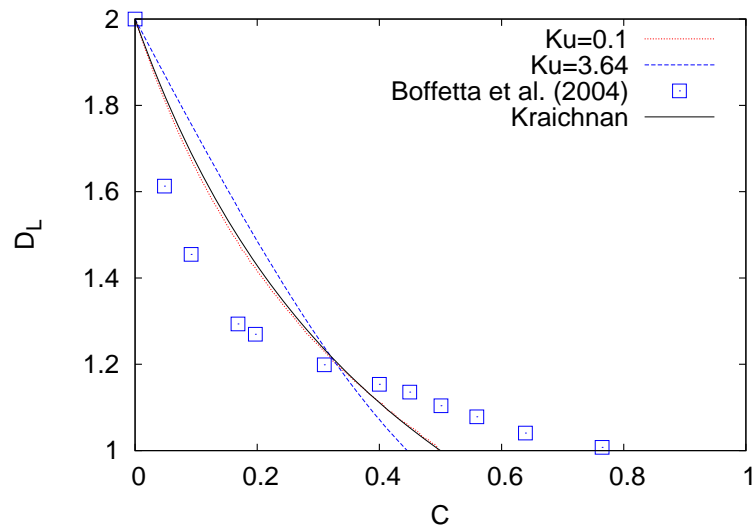}
\caption{(Colour online)
$D_L$ as a function of $ C$ for the Kraichnan model (solid, black curve) and
for the compressible renewing flow with $\Ku=0.1$ (dotted, red curve) and
$\Ku=3.64$ (dashed, blue curve). The square symbols are from the direct numerical simulation
of a turbulent surface flow by \citet{BDES04}.}
\label{fig:comparison}
\end{figure}

The comparison of our results 
for intermediate values of $\Ku$
with those by \citet{BDES04}
reveals yet another difference.
In the compressible renewing flow, 
the clustering is reduced compared to the small-$\Ku$ case if the
compressibility is small, but it is enhanced if the
compressibility is large. 
This behaviour is the opposite of that found in the turbulent surface
flow (figure~\ref{fig:comparison}). Moreover,
in the renewing flow, the critical degree of compressibility is less than or equal to
1/2 for all values of $\Ku$;
in the surface flow, it is significantly greater (figure~\ref{fig:comparison}).
In the light of our findings, 
it would be interesting to examine the distribution of the stagnation
points of the surface flow considered by \cite{BDES04} 
and understand how it influences the statistics of clustering.

In conclusion, the differences between our findings and those obtained in different flows 
demonstrate that the properties of tracer dynamics in time-correlated
compressible flows are strongly non-universal, to the extent that flows with comparable
$C$ and $\Ku$ can have an opposite effect on clustering. 
In particular, the level of
clustering depends dramatically on the peculiar structures
of the velocity field toward which tracers are attracted. 


\bigskip\smallskip
The authors are grateful to B. Mehlig and K. Gustavsson for useful discussions.



\end{document}